\documentclass[a4paper,11pt]{article}
\usepackage{jinstpub} 
\usepackage{upgreek}
\usepackage{url}
\usepackage{adjustbox}
\usepackage[normalem]{ulem}

\newcommand{\todo}[1]{}
\renewcommand{\todo}[1]{{\color{red} TODO: {#1}}}

\title{\boldmath Inter-pixel cross-talk as background to two-photon interference effects in SPAD arrays}

\author[a, 1]{Sergei Kulkov\note{Corresponding author.},}
\author[a]{Tereza Potuckova,}
\author[b]{Ermanno Bernasconi,}
\author[b]{Claudio Bruschini,}
\author[b]{Tommaso Milanese,}
\author[b]{Edoardo Charbon,}
\author[c]{Mst Shamim Ara Shawkat,}
\author[a,c]{Andrei Nomerotski,}
\author[a]{Peter Svihra}

\affiliation[a]{Faculty of Nuclear Sciences and Physical Engineering, Czech Technical University in Prague, 115 19 Czech Republic}
\affiliation[b]{
École polytechnique fédérale de Lausanne (EPFL), CH-2002 Neuchâtel, Switzerland
}
\affiliation[c]{
Department of Electrical and Computer Engineering, Florida International University, Miami FL 33199, USA
}

\emailAdd{sergei.kulkov@fjfi.cvut.cz}

\abstract{Cross-talk is a well-known feature of single photon avalanche detectors. It is especially important to account for this effect in applications involving temporal coincidences of two or more photons registered by the sensor since in this case the cross-talk may mimic the useful signal. In this work, we characterize the cross-talk of the LinoSPAD2 detector, as well as perform joint measurements of the cross-talk and Hanbury Brown - Twiss two-photon interference, comparing and cross-calibrating both effects. With a median dark count rate of 125 cps/pixel, we report the average cross-talk probability of 0.22\% for the nearest neighbor and also observe a long-range cross-talk of the order $2 \cdot 10^{-5}$~\% for channels separated by up to 20 pixels.}

\keywords{Intensity interferometry, Photon detectors for visible photons, CMOS imagers, Solid state detectors, Timing detectors}

\arxivnumber{2406.15323} 

\begin{document}
\maketitle
\flushbottom

\section{Introduction}
\label{sec:intro}

Multichannel single-photon sensitive detectors with good timing resolution (10--100 ps) and photon detection efficiency (PDE) approaching unity, dark count rate (DCR) lower than 100 cps, low cross-talk (<0.1\%) and afterpulsing are very desirable in a variety of research fields such as fluorescence lifetime imaging \cite{lichtman2005fluorescence, charbon2014single, bruschini2019single}, quantum communications and networks \cite{ciurana2014quantum, Ianzano2020}, remote sensing \cite{aasen2018quantitative, Yingwen2020}, and stellar interferometry \cite{gottesman2012longer, Nomerotski_2020, stankus2022two, Crawford2023} to name a few. There are multiple options of single photon sensitive detectors available today, including charge-coupled devices (CCD) \cite{jost1998spatial, brida2008measurement, zhang2009characterization, fickler2013real, avella2016absolute, reichert2017quality, moreau2019imaging}, time-stamping complementary metal-oxide-semiconductor (CMOS) cameras \cite{jachura2015shot, timepixcam, Nomerotski2019, Nomerotski2023}, transition-edge sensors (TES) \cite{cabrera1998detection, lita2008counting}, superconductive nanowire single-photon detectors (SNSPD) \cite{divochiy2008superconducting,natarajan2012superconducting, holzman2019superconducting, zhu2020resolving, korzh2020demonstration}, and single-photon avalanche diodes (SPAD) detectors \cite{charbon2014single, perenzoni2016compact, gasparini2017supertwin, bruschini2019single, morimoto2020megapixel, lubin2021heralded, wojtkiewicz2024review}. To reach the single-photon sensitivity, the CCD cameras require external intensifiers or an internal electron-multiplication stage and typically have low frame rates, while the TES and SNSPD detectors require cryostats at liquid helium temperature and are, therefore, difficult to scale. During the last years, the SPAD sensors had fast progress in improving main operational parameters and currently have notable practical advantages over the other technologies, namely operation in ambient conditions without cooling with picosecond scale timing resolution while providing linear and 2D arrays with a multitude of synchronized channels \cite{gramuglia2021low, gramuglia2022sub}.  The peak photon detection efficiency (PDE) in SPADs is not as good as in SNSPDs but lately there was a considerable progress in improving it to a reasonable 30-50\% \cite{marsili2013detecting, reddy2020superconducting}. As a downside, as many detectors with internal multiplication, SPADs are vulnerable to the afterpulsing \cite{ziarkash2018comparative} and cross-talk \cite{rech2008optical}. 

Minimization of the cross-talk could be critical for all aforementioned research fields as it can act as background, especially when a temporal correlation of two or more single photons serves as a signature for the process under investigation. These signatures can arise in different situations when simultaneous pairs of photons are produced, like in the spontaneous parametric down-conversion in quantum photon sources \cite{SPDC_general, PRL_Oxford_spdc2008, spdc_grice_PRA97}, or in the temporal second-order correlation of photons, like in the Hong-Ou-Mandel (HOM) \cite{HOM_effect, Bouchard_HOM_review, Sensors2020_Nomerotski} and Hanbury Brown-Twiss (HBT) \cite{brown1957interferometry, brown1958interferometry1} effects. In these cases the cross-talk could mimic a coincidence providing the second hit, which would accompany the real photon hit. We note that even a cross-talk with small probability could be a tricky background since the signal coincidence signatures can be rare as well.

Cross-talk is a well-known feature of many semiconductor detectors. There could be multiple cross-talk mechanisms \cite{rech2007depth, rech2008optical, xu2014crosstalk, xu2016cross, ficorella2016crosstalk, kroger2017high, ficorella2017crosstalk, jahromi2018timing, ratti2021cross}, but the primary one is the generation of secondary photons in the electron avalanche during the multiplication. These photons can travel some distance in the sensor and could trigger another avalanche event, thus, producing another signal in a different pixel. In this work, "SPAD", "SPAD cell", and "pixel" are used interchangeably, and in the context of the LinoSPAD2 detector, always refer to one of 512 SPADs.. The cross-talk probability increases for SPADs with smaller cells and with higher density of the  diodes. To overcome this, trenches surrounding cells are used to isolate each SPAD from its neighbors. 
While it was successful in suppressing the direct cross-talk, other mechanisms like the indirect optical paths due to reflections off the edges and other structures in the sensor can keep the cross-talk probability non-zero for pixels separated by considerable distance. We note that similar studies also have been performed for analog silicon photomultipliers (SiPM) \cite{Renker2009, Nakamura2019}.

The HBT effect that was discovered in the late 50s and later used, for example, for measuring diameters of stars \cite{brown1958interferometry2, brown1958interferometry3, hanbury1979test} is essentially the tendency of thermal photons 
to bunch together in time at certain conditions. In terms of coherency, the HBT effect states that thermal photons could be second-order coherent and will result in a $g^{2}(\tau, \Vec{r})$ function, where $\tau$ is a delay between detections of the two photons, and $\Vec{r}$ is the distance between detectors, with a visible peak of $1 < g^{2}(\tau=0) < 2$ \cite{loudon2000quantum}. Measurement of the HBT peak is the cornerstone of the stellar intensity interferometry \cite{walter2023resolving, guerin2017temporal, guerin2018spatial, rivet2020intensity, abeysekara2020demonstration, de2022combined, karl2024photon}. While for the original measurements done by Hanbury Brown and Twiss, single-channel devices were used, such as photomultiplier tubes, where the cross-talk effect does not apply, recent developments in the SPAD technology led to a shift to the multichannel detection systems. In terms of photon coincidences, the HBT and cross-talk effects would look essentially the same: both would produce a visible temporal coincidence peak at the same position of $\tau=0$. While the cross-talk effect is essentially inherent to the SPAD sensor, the HBT effect depends on the phase delays of the photon optical paths before the sensor affecting the resulting position of the HBT peak in the time difference distribution. This can be used to separate the two effects and, thus, to distinguish the useful HBT peaks from the background peaks produced by the cross-talk effect.

\begin{figure}
    \centering
   \includegraphics[width=0.6\linewidth]{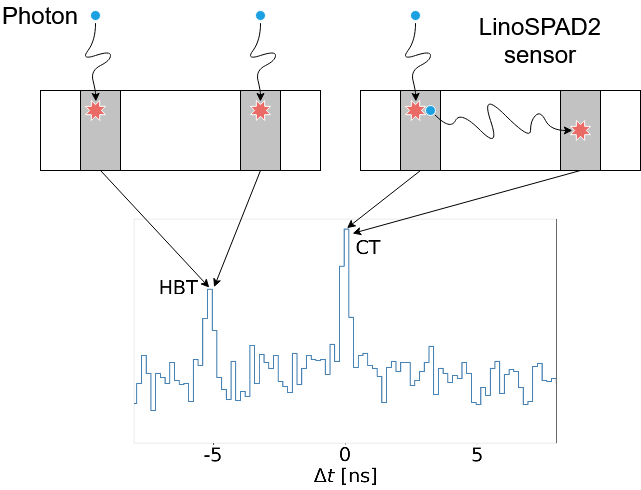}
    \caption{Left: HBT effect mechanism. Detection of two photons is required to observe the effect. Photons with similar wavelengths arriving at the sensor close in time will produce an HBT peak in a coincidence histogram. The position of the peak can be manipulated by extending one of the photon paths.
    Right: Cross-talk mechanism. Detection of only one photon is required to observe the cross-talk effect. Upon generation of the primary electron avalanche, a secondary photon can be generated that may travel to neighboring pixels triggering another avalanche.}
    \label{fig:mechanism}
\end{figure}

Figure \ref{fig:mechanism} shows mechanisms for the cross-talk and HBT effects. For the cross-talk, one signal photon is required that with some probability will generate a signal in another cell of the sensor through the cross-talk mechanism. On the other hand, the HBT effect can be seen only for pairs of photons, hence the registration of two signal photons is required. While both effects produce peaks in the time coincidence distribution that can be constructed comparing timestamps of different photons, the cross-talk effect is inherent to the SPAD sensor, and the position of the cross-talk peak will depend only on the internal delays introduced by the detector electronics. The HBT peak position, however, can be manipulated via additional delays in photon paths before they reach the sensor. 
To differentiate between the two effects, one of the photons has an additional 5 ns delay in its path.

In this work, we aim to characterize the cross-talk effect in the LinoSPAD2 detector with a linear array of SPADs and also to study how it can be compared to the HBT effect measured with the same sensor in identical conditions.

\section{Experimental setup}

\subsection{The LinoSPAD2 detector}

The LinoSPAD2 is a single-photon sensitive detector with a linear array of 512 SPADs on 26.2 $\upmu$m pitch with timing precision of approximately 40 ps (rms) \cite{bruschini2023linospad2, milanese2023linospad2, jirsa2023fast}. Each half of the sensor is read out with a separate Spartan 6 Field-Programmable Gate Array (FPGA). For the sensor without microlenses with a native fill factor of 25.1\%, which was used for these measurements, the PDE covers the whole visible spectrum and reaches the near-infrared, peaking at 13\% for 530 nm. The fill factor is defined as the ratio between the photosensitive and total pixel areas, where the total area may also include associated pixel electronics and trenches. The average cross-talk probability of 0.19\% and the median dark count rate (DCR) of $\sim70$ cps/pixel were measured at $20^{\circ}\mathrm{C}$ and excess bias voltage of 6 V (the breakdown voltage is 24 V) during previous characterization of the sensor \cite{milanese2023linospad2}. 

In this work, the detector was operated in the timestamping mode when both the time of photon arrival and the pixel where the photon was detected were recorded. The data recording was done for sequences of 4 ms cycles with each data file corresponding in total to 64 seconds (16000 cycles). Data transfer from the FPGAs to PCs was performed via USB 3.0 in 32-bit long words, where both the photon timestamp and corresponding pixel number are encoded.

\begin{figure}[htbp]
\centering 
\adjustbox{valign=c}{\includegraphics[width=.37\linewidth]{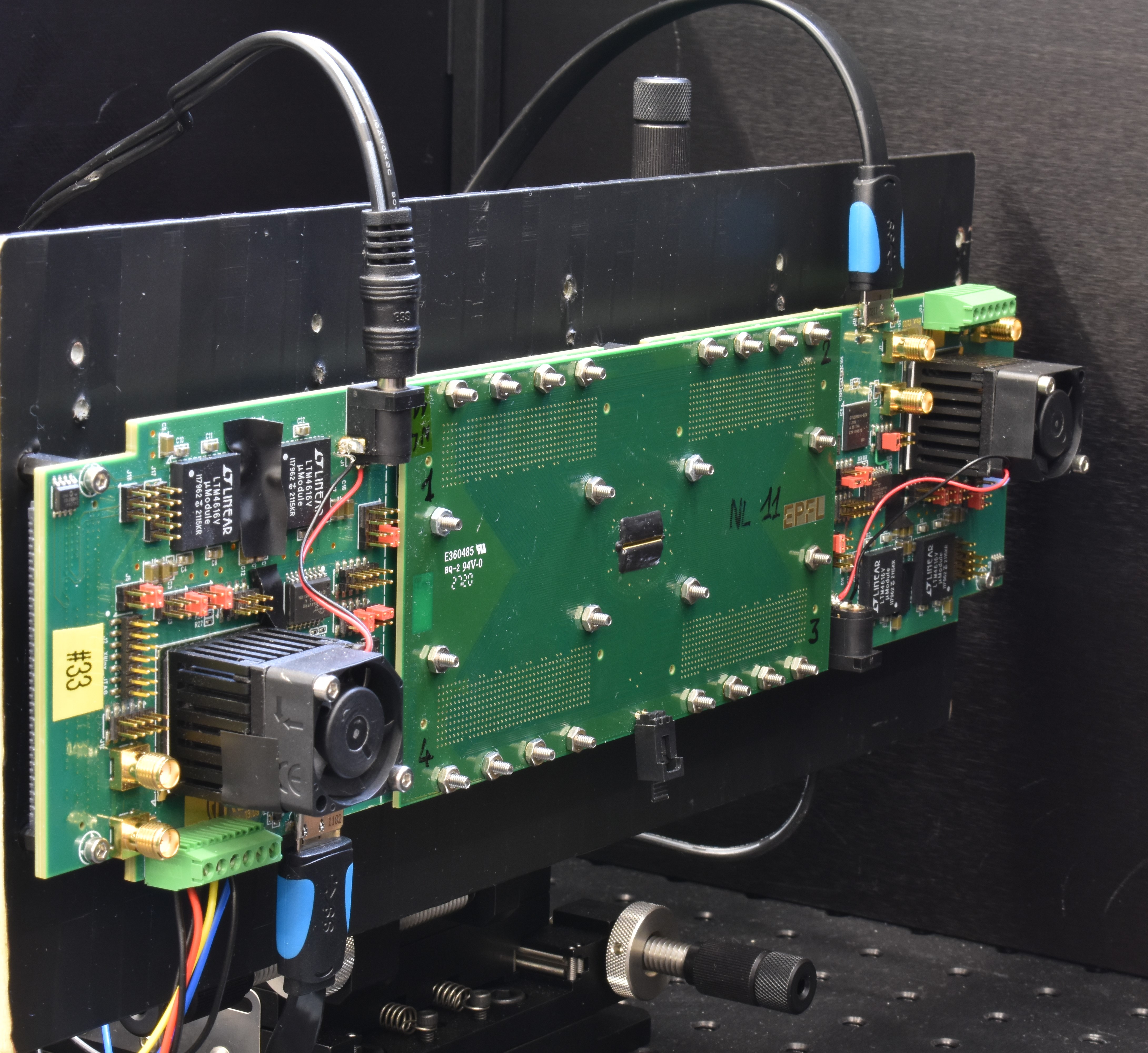}}
\qquad
\adjustbox{valign=c}{\includegraphics[width=.55\linewidth]{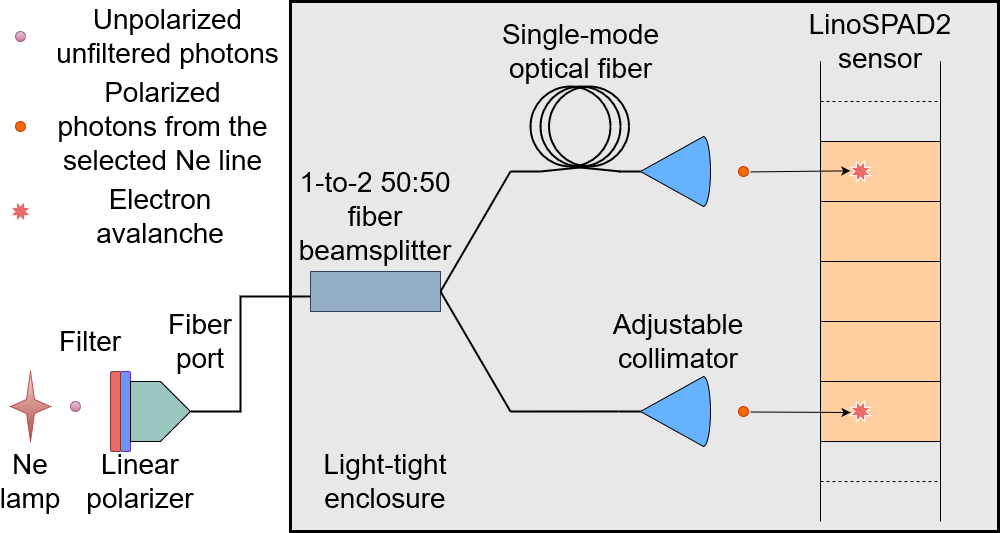}}
\caption{\label{fig:setup_photo} Left: LinoSPAD2 inside a light-tight enclosure. 
Right: Setup diagram. Light from the Ne lamp is polarized, spectrally filtered, and coupled into a 1-to-2 50:50 single-mode fiber beamsplitter. The beamsplitter outputs are connected to two adjustable collimators focusing the light onto the LinoSPAD2 sensor. A single-mode optical fiber of variable length was used to introduce a delay into one of the beamsplitter outputs. Except for the lamp and fiber port with filter and polarizer, all components were placed in a light-tight enclosure.}
\end{figure}

For all measurements, the LinoSPAD2 detector was placed in a light-tight enclosure blocking the ambient light. All other sources of light, like LEDs on the readout board, were also blocked. Figure \ref{fig:setup_photo} shows a photograph of the LinoSPAD2 sensor inside the enclosure together with a simplified diagram of the whole setup.

\begin{figure}[htbp]
\centering 
\adjustbox{valign=c}{\includegraphics[width=.45\textwidth]{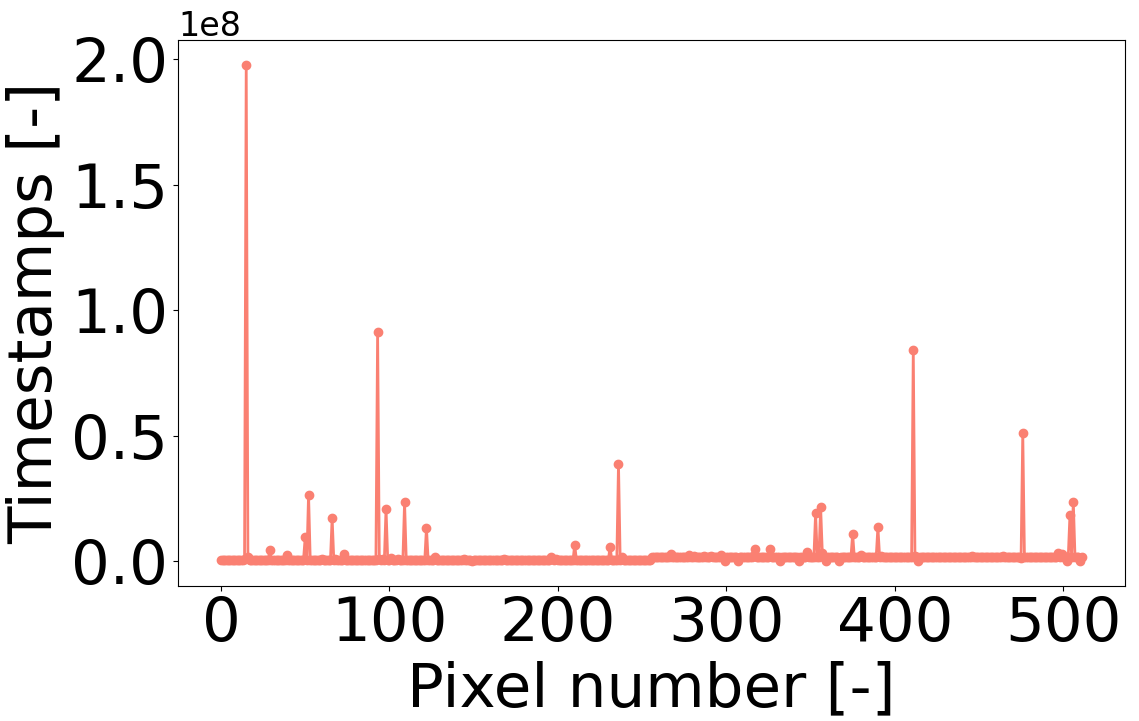}}
\qquad
\adjustbox{valign=c}{\includegraphics[width=.45\textwidth]{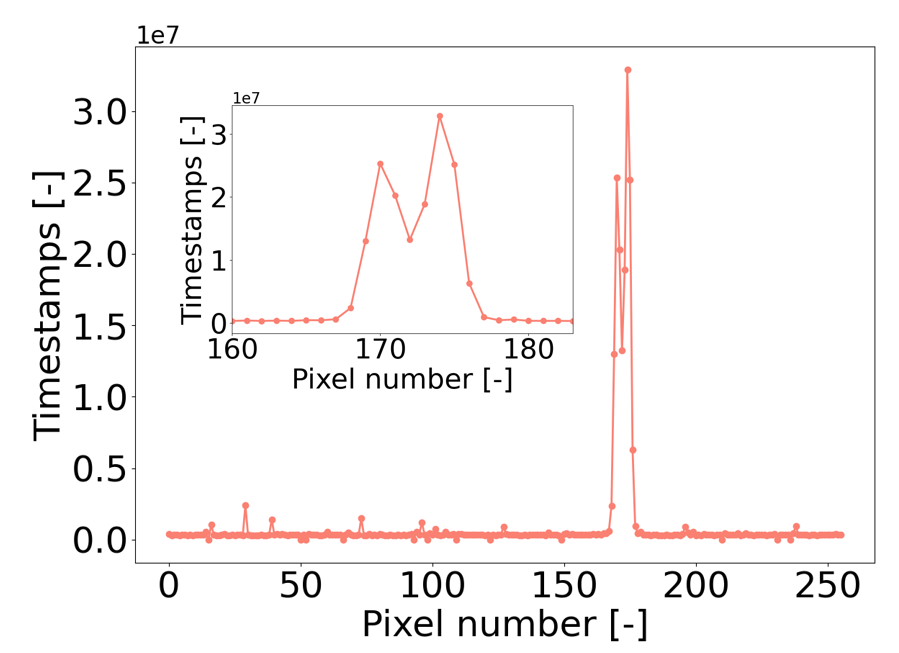}}
\caption{\label{fig:CT_HBT_senpop} Left: Sensor occupancy from data used for cross-talk measurements showing hot pixels with high DCR from the full sensor. Right: Sensor occupancy from data used for the joint CT/HBT measurements with two Ne lamp outputs on the sensor. Only one half of the sensor was used for the CT/HBT measurements.}
\end{figure}

For the joint cross-talk and HBT measurement, the detector was put in the same light-tight enclosure but this time a thermal source of light was used to measure the HBT effect. For this, a Ne calibration lamp (Newport 6032) powered with 10 mA power source (Newport 6045) was used. The Ne spectrum has well-defined and separated lines, which can be easily detected with LinoSPAD2. The light was collected into a single-mode wideband coupler used as a 50:50 beamsplitter (Thorlabs TW670R5F1), where the input was connected to a fiber port (Thorlabs PAF2-5A) at the source, and two outputs were connected to two adjustable collimators (Thorlabs CFC8A-A) that were used to focus the light onto the sensor. This resulted in two illuminated areas at the sensor, each of about 5 pixels in diameter, see Fig. \ref{fig:CT_HBT_senpop}. 

The highest photon rate was registered by pixel 174 at 31 kHz. The rate was determined from the total number of recorded timestamps divided by total duration of recorded cycles without accounting for possible delays in communications over the USB line, which could vary but were never considerable. The difference in photon rates seen by the two pixels with the highest rate can be explained by the unequal split of the 50:50 beamsplitter. As the cross-talk probability is expected to decrease rapidly with the distance between the aggressor (i.e. emitter of cross-talk) and victim (i.e. the detector) pixels, placing the two photon beams with minimal distance between them should result in the visible cross-talk signals. On the other hand, if photons satisfy the coherence conditions for the HBT effect --- namely being of the same wavelength and arriving at the detector close in time --- the HBT peak will be visible in the detector for any pair of hit pixels. In our setup, the same spatial mode is ensured by coupling the lamp light to a single-mode fiber and employing the single-mode fiber-coupled beamsplitter to form two beams hitting the SPAD array. The same frequency is garanteed by using a single line in the neon spectrum.

\subsection{LinoSPAD2 calibration}

There are two main calibrations that can be applied to the LinoSPAD2 detector. One of them is to account for differences in the width of  TDC (time-to-digital converter) bins in the FPGA. Nominally, each bin covers 17.857 ps, but in reality, due to process variations, each bin may have a different duration. The actual width of each TDC bin can be estimated by collecting a data set with the whole sensor homogeneously illuminated. There are 140 bins in each TDC and by plotting a histogram of the raw binary data with 140 bins, one can estimate the width of each bin from the resulting statistics and use it later in the data analysis. This calibration procedure is mandatory as it directly affects the precision of all timing measurements.

The second calibration accounts for different delays of electrical paths between the sensor SPADs and corresponding TDCs in FPGA. A small delay introduced by these paths is added to each photon timestamp and that delay will be different for each SPAD cell. While it does not affect the precision of the measurements for a pair of single pixels, it results in a shift of the resulting cross-talk and HBT peaks relative to $\Delta t =0$ in the coincidence plots. Therefore, this calibration procedure is not mandatory and it is not applied in this work as we always used a single pair of pixels for the measurements. We note that if the information from multiple pixels was to be combined then this correction would need to be applied to account for different time offsets in the pixels.

\subsection{Theoretical models for cross-talk and HBT}

To extract the cross-talk probability from the data, pixels with high DCR were chosen as aggressor pixels. To see how the cross-talk probability changes with the distance from the aggressor pixel, 20 neighbouring pixels on each side are used as victim pixels, and where not applicable, like on the edges of the sensor, neighbours only from one side are used. The cross-talk hits have timestamps with minimal time difference relative to the primary photon. Therefore, the distribution of timestamp differences will exhibit a cross-talk peak. This peak can be fitted using a Gaussian function to extract the peak position and standard deviation (sigma), with the latter corresponding to the detector timing jitter. The sigma can also be used to calculate the counts in the peak $N_{\mathrm{CT}}$ by taking all data points in a $\pm2\sigma$ interval around the peak position corresponding to 95\% of the counts under the peak. Background $N_{\mathrm{bckg}}$, which is present in these counts, should be estimated and subtracted employing another $\pm2\sigma$ interval in a region far away from the peak. This procedure is done for each pair of pixels where the first pixel is the aggressor one and the second one is from the 20 victims to each side of the aggressor. To calculate the cross-talk probability between the two given pixels with distance $d$ in pixels between them, the peak population should be divided by number of photons registered in each pixel as both pixels contribute to the cross-talk effect:

\begin{equation}
    P_{\mathrm{CT}, d} = (N_{\mathrm{CT}} - N_{\mathrm{bckg}}) / (I_{1} + I_{2}) \cdot 100\%,
    \label{eq:CT_prob}
\end{equation}
where $I_{1}, I_{2}$ is the light intensity seen by the two pixels given as the number of photons detected in the given amount of time.

The HBT effect states that photons coming from a thermal light source have temporal bunching when certain coherence conditions are met. Therefore, by looking at the timestamp differences of thermal photons, one should see an HBT peak on top of a flat background of random coincidences. The peak can be again fitted using the Gaussian function to extract the peak position, height of the peak that gives the contrast, or the ratio of the peak height to the background level of random coincidences, of the HBT peak, and sigma that corresponds to the measured source coherence time \cite{karl2024photon}: 

\begin{equation}
    \tau_{\mathrm{c}} = \frac{1}{\Delta\nu}=\frac{\lambda^{2}}{c\Delta\lambda},
    \label{eq:coh_time}
\end{equation}
where $\Delta\nu$ is the source spectral width, $c$ is the speed of light, $\lambda$ is the photon wavelength, and $\Delta\lambda$ is the source spectral width. For any source of light with a broadband spectrum, the coherence time will be very low. To improve that, spectral filters are usually utilized \cite{guerin2017temporal, karl2024photon}. Here, we use a filter with a central wavelength of 700 nm and FWHM (Full Width at Half Maximum) of 10 nm (Thorlabs FBH700-10) to select the 703.24 nm Ne spectral line. The width of the chosen filter  is rather large 
and, therefore, should result in a coherence time lower than a ps. However, as will be shown in the next section, the actual coherence time we measure is larger. The reason is that Ne spectral lines are much narrower than the filter spectral width, and in our setup, the filter is only used to select a single Ne line. For the detector resolution that is comparable to the source coherence time $\tau_{\mathrm{det}} \approx \tau_{\mathrm{c}}$, the expected maximum contrast can be estimated as \cite{loudon2000quantum}:

\begin{equation}
    C \approx \frac{1}{2}\frac{\tau_{\mathrm{c}}^{2}}{2\tau_{\mathrm{det}}^{2}}\left[\exp\left(-2\frac{\tau_{\mathrm{det}}}{\tau_{\mathrm{c}}}\right) - 1 + \left(2\frac{\tau_{\mathrm{det}}}{\tau_{\mathrm{c}}}\right)\right],
    \label{eq:HBT_C}
\end{equation}
where $\frac{1}{2}$ is due to the unpolarized nature of the measured light as photons need to have the same polarization to interfere. To improve contrast at the cost of two-fold decrease in photon rates, we utilize a linear polarizer (Thorlabs LPNIRE100-B) to select a single polarization. 

Unlike the case of the cross-talk signal, the position of the HBT peak relies on the arrival times of two signal photons. Therefore, if one of the two photons is artificially delayed by increasing the distance of that photon from the source to the sensor, the resulting HBT peak will be shifted accordingly. This technique can be used to distinguish between the cross-talk and HBT peaks. To achieve that, we inserted a single-mode optical fiber (Thorlabs P1-630Y-FC-1) of variable length to one of the beamsplitter arms, delaying one of the photons and thus shifting the HBT peak.

As the cross-talk effect relies on only a single photon arriving at the detector, its probability is linearly dependent on the light intensity, see Eq.~\eqref{eq:CT_prob}. HBT, on the other hand, is a correlation effect between two photons, and, therefore, is dependent on the product of the intensities of both photon beams:
\begin{equation}
   N_{\mathrm{HBT}}  \sim  I_{1}I_{2}.
   \label{eq:HBT_quad}
\end{equation}
We note that this model is a simplification as it does not account for the spatial and temporal modes of the photons.

\section{Results and discussion}

\subsection{Cross-talk measurement}

\begin{figure}[htbp]
\centering 
\adjustbox{valign=b}{\includegraphics[width=.49\textwidth]{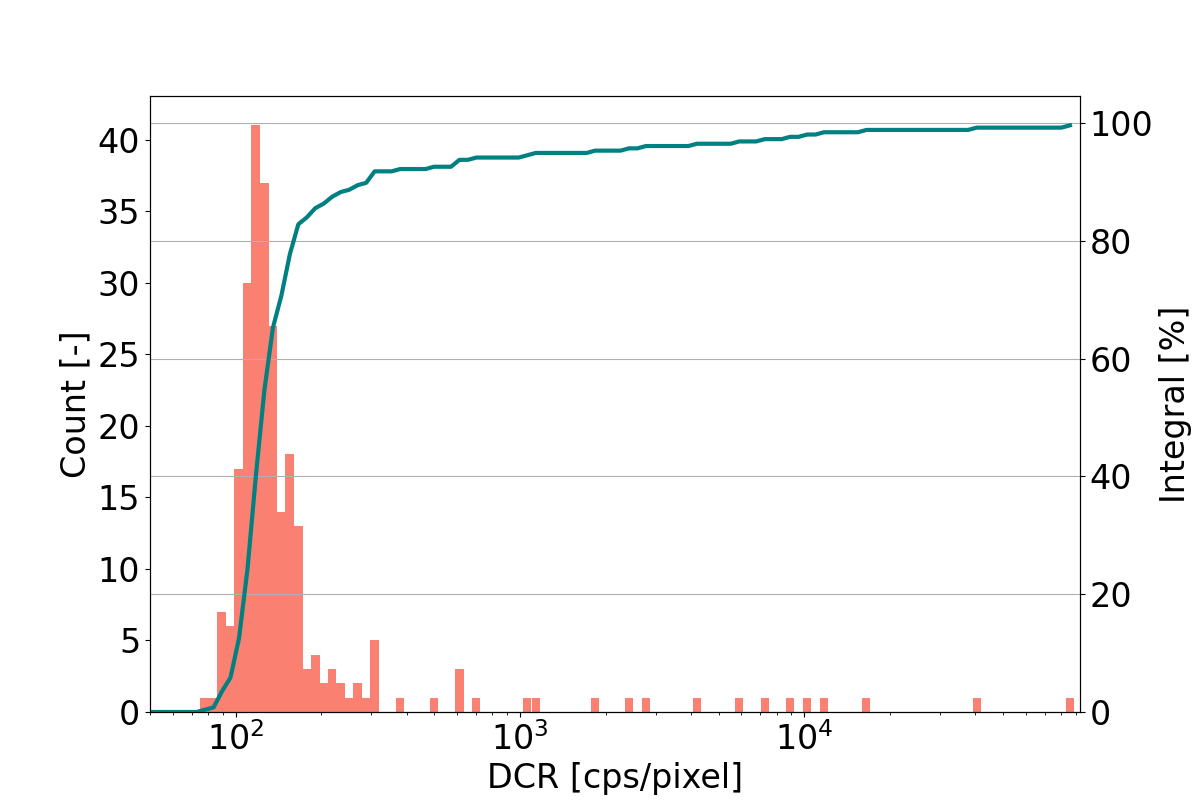}}
\qquad
\adjustbox{valign=b}{\includegraphics[width=.45\textwidth]{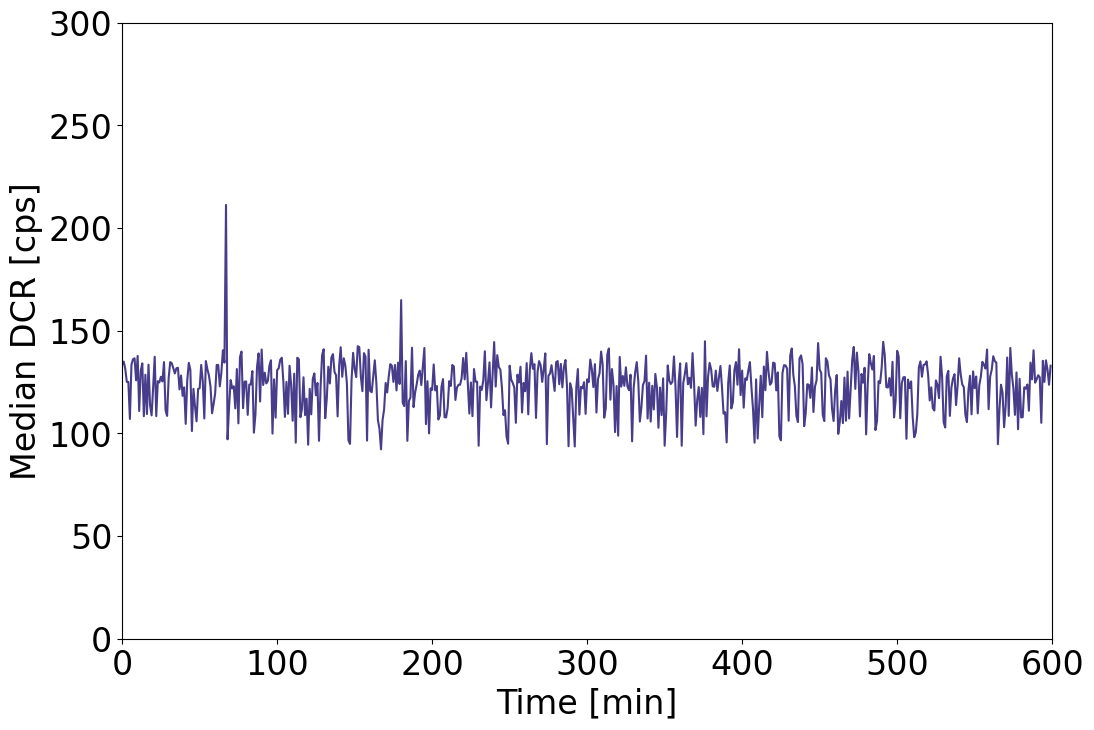}}
\caption{\label{fig:DCR_plots} Left: Distribution of pixel DCRs and its integral for one of the sensor halves. Fourteen pixels (5\% of total) have DCR above $10^{3}$ cps. The median DCR is 125 cps/pixel.
Right: Evolution of median DCR of the same sensor half with time. Except for a couple of occasional peaks, the DCR was stable during $\sim10$ hours of data collection.}
\end{figure}

Data for the cross-talk evaluation was collected for $\sim 40$ min. While the median DCR was estimated at 125 cps/pixel, the highest DCR of hot pixels reached $8.6\times10^{4}$ cps. The total DCR per sensor half of 256 pixels takes about 1.5\% of total readout throughput in our configuration, so is not significant. It is also possible to mask individual pixels, and by masking the 14 pixels with the highest DCR, it is possible to reduce the DCR to the total throughput ratio of 0.2\%. The left plot in Fig. \ref{fig:DCR_plots} shows the distribution of pixel DCRs and its integral for one of the sensor halves. Fourteen pixels (5\% of total pixel count) have DCR above $10^{3}$ cps. The other half of the detector behaved similarly. It is also important to check the stability of DCR as it may change with time due to temperature fluctuations. The right plot in Fig. \ref{fig:DCR_plots} shows how the median DCR of the same sensor half changes with time for the whole duration of the data collection. Except for a couple of spikes, which can be attributed to manipulations around the setup that led to stray light reaching the sensor, the median DCR was stable.

\begin{figure}
    \centering
    \includegraphics[width = 0.65\linewidth]{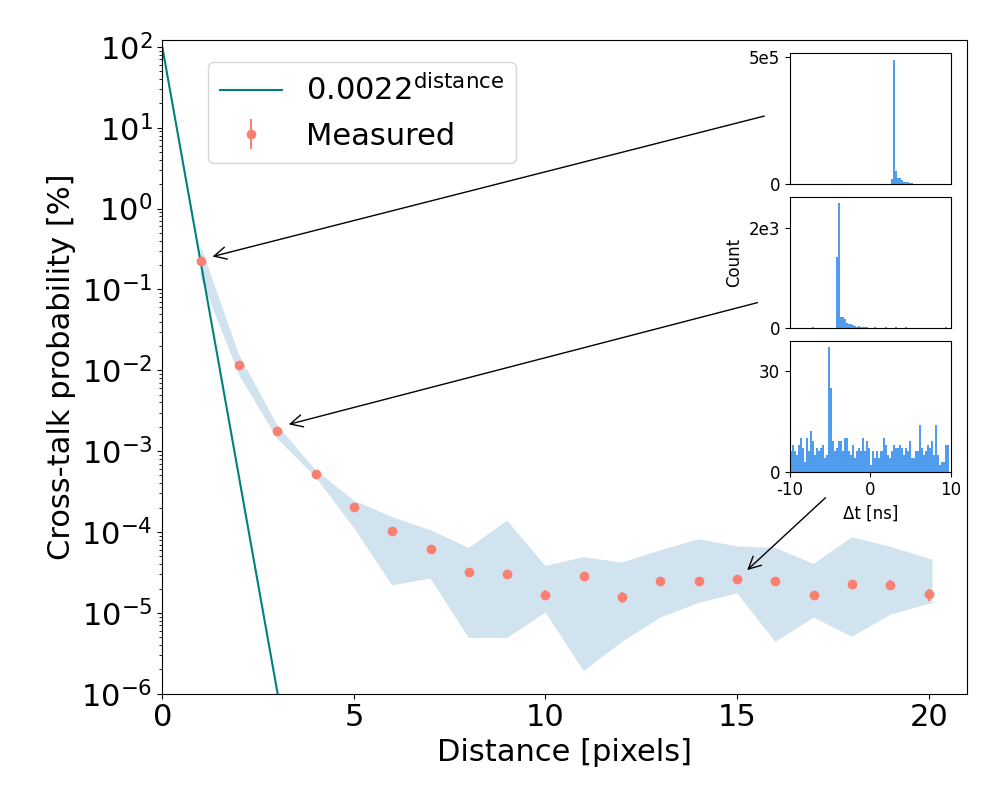}
    \caption{Average cross-talk probability as a function of distance from the aggressor pixel.
    The simplest model for the expected cross-talk probability, when it follows an exponential decrease with the distance from the aggressor, is shown in green. The inserts show examples of the photon coincidence histograms where CT peaks are visible for different pairs of pixels. The tails seen in the histogram peaks for pixel distances 1 and 3 is most probably due to cross-talk from the neighboring pixels. The shaded area around the data points represents the spread of CT probabilities from the minimal to the maximum measured for those distances.}
    \label{fig:Average_CT}
\end{figure}

The average cross-talk probability dependence on the distance between the aggressor and the victim pixels can be seen in Fig. \ref{fig:Average_CT}. All pixels with DCR higher than 4000 cps were used to produce this plot. The simplest model of cross-talk probability was used:
\begin{equation}
P_{\mathrm{CT}} = 0.0022^{d}\cdot100\%,
\end{equation}
where 0.22\% is the average cross-talk probability between the two closest neighbouring pixels, and $d$ is the distance between the aggressor and victim pixels. This model assumes that the cross-talk in the pixel is caused only by the immediate neighbouring pixel.
In reality, there are multiple cross-talk mechanisms at play that are impossible to distinguish in our measurements but nevertheless are evidently present in LinoSPAD2, as seen in the higher long-range cross-talk probabilities. The inserts in Fig. \ref{fig:Average_CT} show the photon coincidence histograms for different pairs of pixels with CT peaks visible. The offset calibration is not applied, therefore the peaks appear shifted relative to $\Delta t =0$. Additionally, for the histograms for pixel distances of 1 and 3, there is a visible tail to the right of the peak. This can be attributed to the cross-talk effects from the neighbouring pixels of the two analyzed. 

While the probability of seeing cross-talk between two distant pixels is very small, of the order $10^{-4} - 10^{-5} \%$, the cross-talk peaks may still appear and be visible. This is especially undesirable when measuring the HBT effect as the two peaks will appear at the same position in the histogram if no delays are introduced to either of the beamsplitter outputs. Therefore, it is preferable to separate the pixels of interest for the HBT measurements by at least 20 pixels to minimize the probability of seeing CT peaks.

\subsection{Joint cross-talk and HBT measurements}

\begin{figure}
    \centering
    \includegraphics[width=0.85\linewidth]{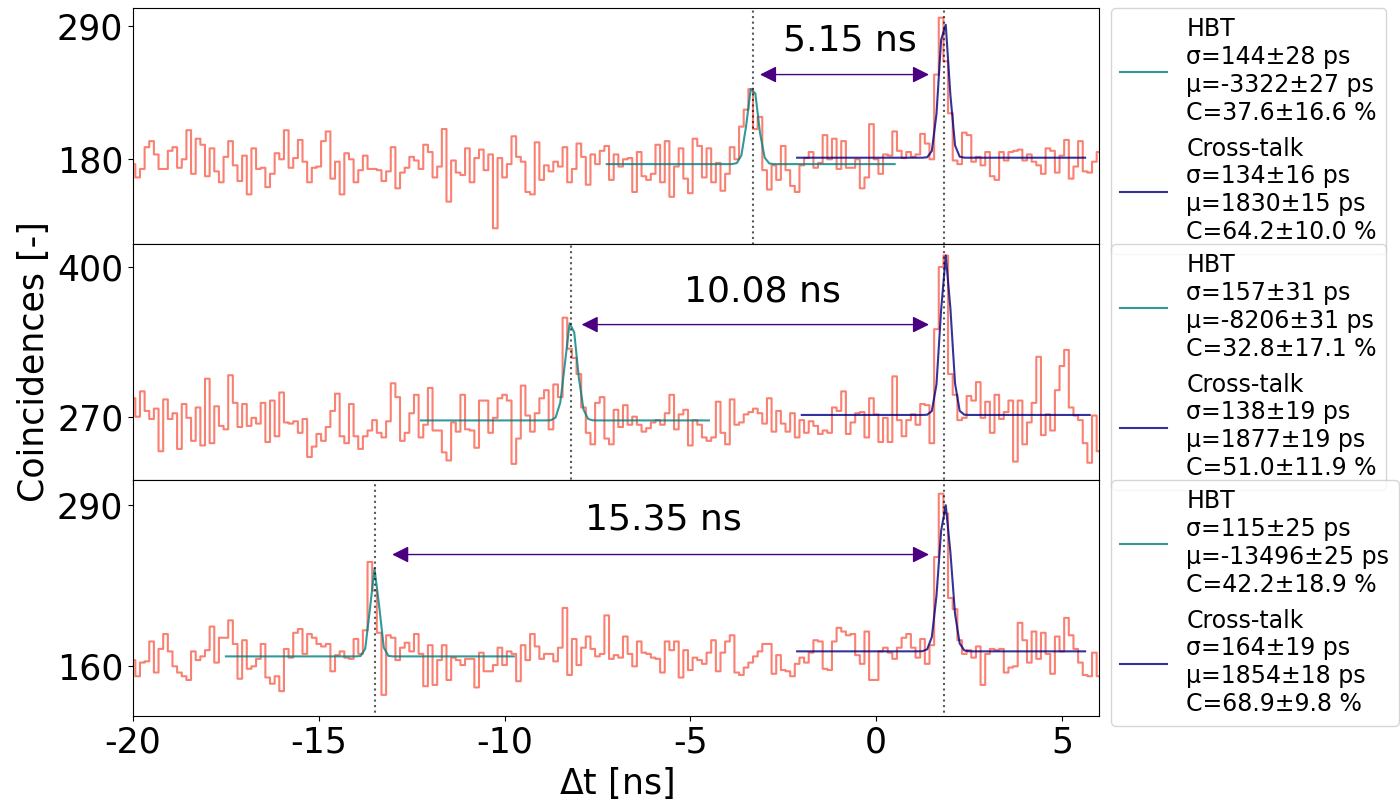}
    \caption{Histograms showing photon coincidences for three different delays introduced to one of the beamsplitter outputs: single-mode optical fiber with a length of 1 m (top), 2 m (middle), and combination of the two resulting in 3 m delay (bottom). The cross-talk and HBT peaks are fitted using the Gaussian function, the fit parameters can be seen to the right of the plots. The arrows show the measured shift of the HBT peak relative to the cross-talk one.}
    \label{fig:CT_HBT_coin_plot}
\end{figure}

To see both the cross-talk and the HBT effects at the same time, the light from the Ne lamp was focused at pixels 170 and 174, see Fig. \ref{fig:CT_HBT_senpop}. As each illuminated dot has a diameter of 5 pixels, the two peaks are partially overlayed. Both outputs of the beamsplitter were positioned at approximately the same distance from the sensor, so without any additional delays in any arm of the beamsplitter, the HBT peak would appear at the same position in the timestamp difference distribution as the cross-talk peak. To shift the HBT peak, a single-mode optical fiber was inserted into one of the beamsplitter arms with a length of 1, 2, and 3 m for a delay of approx. 5, 10, and 15 ns respectively. Only two fibers of length 1 and 2 meters were used, and the combination of the two with the use of a FC/PC-to-FC/PC adapter resulted in a fiber 3 meters long. Figure \ref{fig:CT_HBT_coin_plot} shows both the CT and HBT effects as they are seen in the photon coincidence plot for three different data set with three different delays introduced. The CT peak is not located at $\Delta t=0$ because the offset calibration is not applied, and the HBT peak is shifted relative to the CT one due to an additional single-mode optical fiber introduced to one arm of the beamsplitter. Both peaks are fitted with the Gaussian function to extract their positions and the standard deviation, which was used to calculate the population of each peak. The data acquisition time is the same across all three data sets. However, for the data sets with 5 ns and 15 ns delays, the number of coincidences is lower due to the 1 m-long single-mode fiber, which most probably has a dusted core and thus leads to a loss of photons and which was utilized in both measurements. We note that for the HBT effect, both photons may travel through the same beamsplitter arm and thus reach the same group of the 5 pixels where the output is focused. Therefore, the HBT peak for such pairs of photons will appear at the same position as the CT peak. However, the contribution of this HBT peak to the overall CT counts under the peak will be small mainly due to much lower intensity at the edges of the focused beam.

The standard deviation for the HBT peak was estimated to be of the order of 150 ps on average. The average contrast of the HBT peak was estimated at $(36 \pm 13)$\% that was averaged over all 3 data sets at full intensity, when the statistics are the highest and, hence, have the best fit quality. However, the highest contrast can be estimated using Eq.~\eqref{eq:HBT_C} at 84\%. 
A possible explanation is that some part of the contrast is lost due to imperfections in the optical components utilized.

With a separation of 3 pixels between the two pixels of interest, the CT effect is still stronger resulting in higher counts under the peak compared to the HBT one, since the average cross-talk probability for that distance is of the order of $10^{-4}$\%. Moreover, the expected HBT contrast in this setup is the highest thanks to the narrow lines of the Ne spectrum. In measurements where the source coherence time is dictated by the filter width or spectrometer binning, such as in setups with broadband sources of light, the CT effect would dominate over the HBT effect. The fact that the width of the CT peak is much larger than the detector timing precision hints at variations of CT photon path delays, possibly due to reflections off the sensor edges. 

\begin{figure}
    \centering
    \includegraphics[width=.6\textwidth]{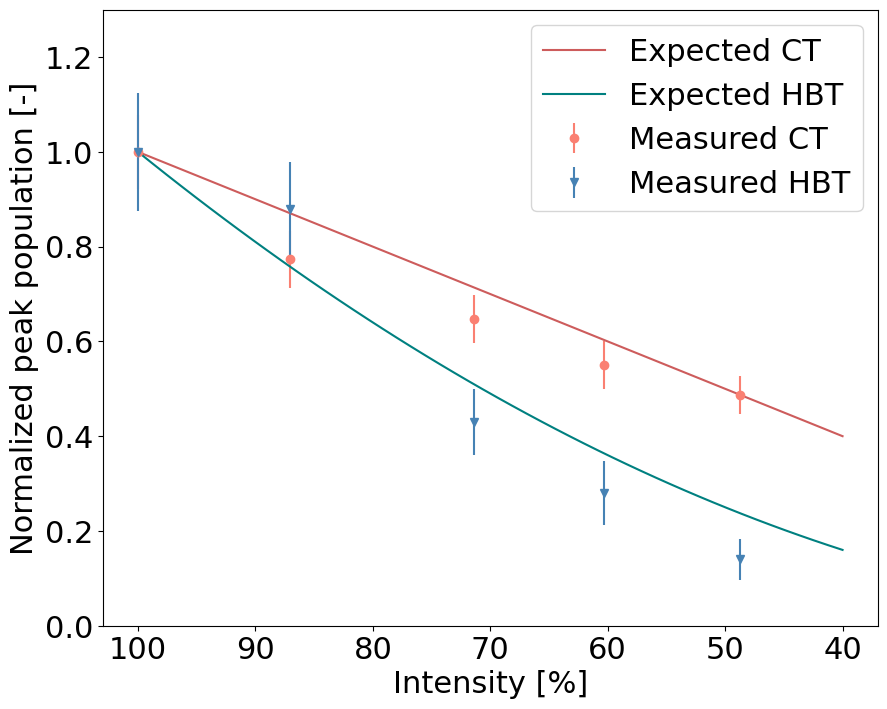}
    \caption{Normalized counts under the peak of the cross-talk and HBT effects as a function of average intensity seen by the pixels 170 and 174 averaged over data sets with 5, 10, and 15 ns delays. The expected trends for both effects are shown as lines: linear dependence for the cross-talk effect, and quadratic dependence for HBT.}
    \label{fig:CT_vs_HBT}
\end{figure}

To see how the counts under the peak of both effects changes with the light intensity seen at each pixel, we shift the lens in the fiber port near the lamp vertically. This way we reduce the photon rates equally at both beamsplitter outputs. The expected values of the counts under the peak for both effects starting from the measured ones at full intensity are estimated using \eqref{eq:CT_prob} and \eqref{eq:HBT_quad}. Figure \ref{fig:CT_vs_HBT} shows the average measured values of CT and HBT counts under the peak for three data sets with 5, 10, and 15 ns delays together with the expected trends for both effects. Since the intensity was controlled manually by changing the opening in the fiber port, the measured intensity was different for each data set. Therefore, averaging was done not only over the measured counts under the peak (y-axis), but also over the measured intensities as well (x-axis). While both effects do not exactly follow the trends, the CT counts under the peak are overall higher than the HBT ones, as was expected. The precision of the measurement can be improved by collecting larger datasets. However, it is hard to predict the stability of the setup since different delays involve manipulations with the setup and hours-long measurements.

\section{Conclusion}

The HBT effect, caused by two-photon interference, is a key methodology for many applications. In this work, we investigated the HBT photon temporal bunching in the LinoSPAD2 sensor by detecting pairs of single photons in two different pixels of the sensor. We also characterized the performance of the LinoSPAD2 detector with emphasis on studies of the inter-pixel cross-talk, which can mimic the HBT signature, producing a peak in the time difference distribution for the pixel pairs with hits. 

The median DCR of the sensor and average cross-talk probability between two neighboring pixels were measured to be equal to  125 cps/pixel and 0.22\%, respectively. The slightly higher DCR, compared to previous measurements, could be attributed to a higher temperature inside the dark enclosure and, possibly, to a small amount of stray light reaching the sensor.
More importantly, we found that the cross-talk effect has a long-range component so the temporal correlation peaks can be seen for pairs of pixels with considerable separation between them, up to 20 pixels, even if the cross-talk probability for such cases is much lower. 
This could be critical for two-photon coincidence measurements, where the cross-talk effect essentially mimics the HBT effect appearing as a peak at the same position, near the time difference close to zero. While the cross-talk probability was estimated from the pixels with high DCR, cross-referencing those measurements with joint HBT/CT ones showed strong agreement, so we conclude that the cross-talk probabilities are applicable for real photons as well.
As a precaution, it could be beneficial to delay one of the photons in the two-photon measurements to ensure the separation between the two types of peaks if both appear in the resulting coincidence plot. 

The cross-talk and HBT effect dependence on the light intensity was measured,
it was shown that both effects follow the expected intensity scaling, linear for the cross-talk and quadratic for the HBT, and so that the HBT effect indeed has a stronger dependence on the light intensity. 
The cross-talk could be an important source of background to HBT studies in fast spectrometers, which can be used for experiments involving broadband light sources and spectral binning \cite{Farella2024, jirsa2023fast, ferrantini2024multifrequencyresolved, iso2024capturing}.

We also note that when switching to a broadband source of light, for example to a halogen lamp, wide-band LED, or starlight, one would have the source coherence time defined by the used spectral filter width or, alternatively, by the spectral binning in the spectrometer. As a numerical example, a 10 nm filter will result in a coherence time of 0.16 ps, according to Eq.~\eqref{eq:coh_time} and the HBT peak contrast of only 0.4\%, emphasising again the importance of the cross-talk background mitigation. This also means that the best possible detector timing and spectral resolutions are essential to see the HBT effect with a broadband source of light.

\acknowledgments

This work was supported by the European Regional Development Fund-Project ``Center of Advanced Applied Science" No. CZ.02.1.01/0.0/0.0/16-019/0000778 and by the Grant Agency of the Czech Technical University in Prague, grant No. SGS24/063/OHK4/1T/14. This work was also supported by the EPFL internal IMAGING project ``High-speed multimodal super-resolution microscopy with SPAD arrays", and by the U.S. Department of Energy, Office of Science, Office of High Energy Physics under Award Number DE-SC0024670.


\bibliographystyle{ieeetr}
\bibliography{bilbio}

\end{document}